\documentclass[12pt]{iopart}

\usepackage{graphicx}
\begin{document}

\title{Counterportation and the two-state vector formalism}

\author{Justin Dressel$^{2,3}$, Gregory Reznik$^1$ and Lev Vaidman$^{1,2}$}

\address {$^1$Raymond and Beverly Sackler School of Physics and Astronomy, Tel-Aviv University, Tel-Aviv 69978, Israel}
\address{$^2$Institute for Quantum Studies, Chapman University, Orange, CA 92866}
\address{$^3$Schmid College of Science and Technology, Chapman University, Orange, CA 92866}

\vspace{10pt}
\begin{abstract}
Hatim Salih discovered a method for transferring a quantum state with no particles present in the transmission channel, which he named counterportation. Recently [H. Salih,  Quantum Sci. Technol. 8, 025016 (2023)], he presented a feasible procedure for its implementation. The modification of the protocol by Aharonov and Vaidman, adopted by Salih, justifies the claim that no photons were present in the transmission channel during counterportation. We argue, however, that there is an error in this paper. The analysis of a simplified protocol, which questions the validity of the two-state vector formalism description of the photon presence in the communication channel, is incorrect.
\end{abstract}

\section{Introduction}

Salih suggested a protocol of counterportation \cite{Salih16}
extending ideas started from interaction-free measurements \cite{IFM}, see \cite{Ho06,Salih13,Li15}. In the past, We criticized these proposals \cite{V07,V14,V16C} (see replies \cite{V14R,V16R}) claiming that in all of them the particle leaves a trace in the communication channel not less than in a location about which there is a consensus regarding particle's presence, and therefore, the name counterfactual is not appropriate. However, Aharonov and Vaidman \cite{AV19} have found a way to modify these proposals to make them indeed counterfactual. In a recent paper Salih proposed a very nice practical implementation of his original idea, see  Fig.~2c of \cite{Salih22}, describing in the text the required modification \cite{AV19}. The current article corrects the analysis of Salih's simplified protocol described in  Fig.~1 (see also Fig.1 of \cite{Salih22}). We argue that the protocol presented in this figure is not counterfactual. Fortunately, Salih's claim that it is equivalent to his main protocol described in Fig.~2c of \cite{Salih22}, is incorrect too, so his main results  are unaffected.

\begin{figure} [ht]
   \begin{center}
     \includegraphics[width=13.0cm]{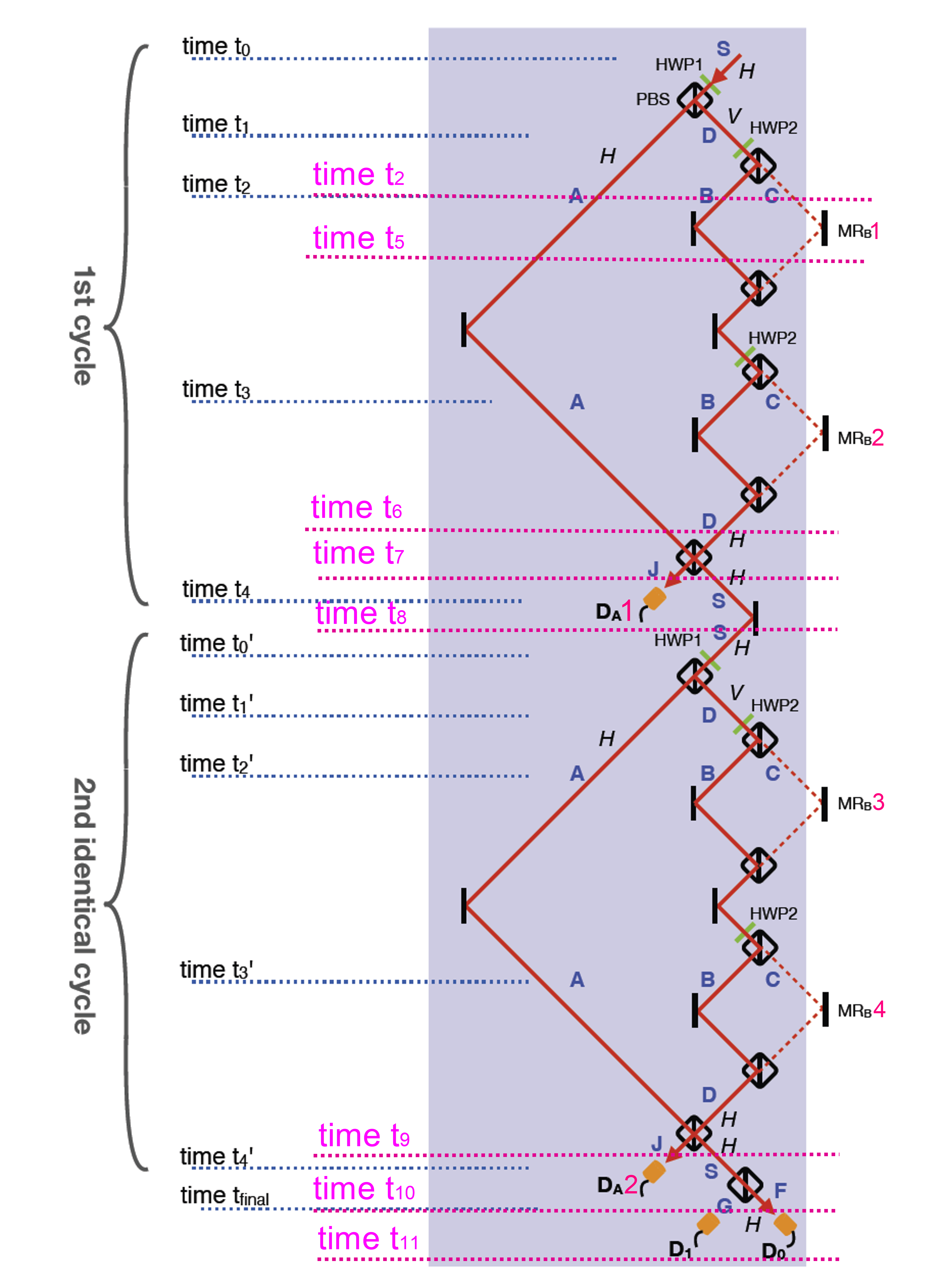}
     \end{center}
   \caption
   { 
   {\bf Salih's interferometric setup}. (Fig.~1 of \cite{Salih22} with additional time points and modified notation.) 
The photon starts at the top, $H$-polarised.  All beam-splitters are polarising, and all half-wave plates (HWP’s) rotate polarisation by 45 degrees. The
setup is such that any photon entering the inner interferometers ends up at detectors $D_Ai$. Salih's paradox is that the two cycles looks identical, but the TSVF asserts that  the photon detected by $D_0$ leaves a trace on mirror $MR_B1$, while it does not leave a trace on mirror $MR_B3$.
}
   \end{figure}

Correcting  Salih's error  removes the question mark about the consistency of the two-state vector formalism (TSVF) analysis of photons inside interferometers \cite{past}. Salih claims that in the setup of Fig.~1 the TSVF suggests that
 ``The weak value [of projection on arm C] is nonzero and consequently a weak measurement in arm
C at time $t_2$ is also nonzero'',  while according to his  ``resolution of the paradox ... the
weak value is thus zero ...'' Or, quoting the caption of Salih's Fig.~1: ``...
consideration of weak measurements of the path
observable at arms C can give paradoxical answers. Our resolution of the paradox, supported
by recent weak-measurement {\it experimental results} \cite{SalihNPJ}, shows that the photon has not been to
any of the right-hand-side mirrors.''

\section{Salih's Paradox}

Salih considers an optical system, see Fig.~1 which he names ``Two outer interferometers, nested within each are two inner interferometers.''   The question he asks: ``Whether a photon detected at detector $D_0$ at the bottom was in any of the arms labeled $C$ leading to the mirrors $MR_B$ on the right-hand side?'' The TSVF analysis tells that the photon was near the  mirror $MR_B1$ (We added the mirror numbers to the original figure) and not near any other mirror $MR_Bi$. The meaning of ``being near the mirror'' is that the photon left a trace on the mirror with magnitude of the same order as a localised photon bouncing off the mirror would leave. Salih finds contradiction with the fact predicted by the TSVF that the photon is near   $MR_B1$ but not near $MR_B3$:
``Yet,
in the absence of a weak measurement, the first outer cycle and the second outer cycle are
identical as far as standard quantum mechanics is concerned—the photon undergoes the same
transformations in each cycle, starting and finishing each in the same state.''

Salih's first analysis within the TSVF is correct. At time $t_2$ the photon is described by the two-state vector $\langle \varphi(t_2) |~~|\psi(t_2)\rangle$, where 
\begin{equation}\label{psi}
 | \psi (t_2)\rangle=\frac{1}{\sqrt 2} |{\rm A},H\rangle + \frac{1}{2} |{\rm B},V\rangle-\frac{1}{2} |{\rm C},H\rangle, 
\end{equation}
\begin{equation}\label{phi}
 \langle \varphi (t_2)| =\frac{1}{\sqrt 2} \langle{\rm A},H| - \frac{1}{2} \langle{\rm B},V|-\frac{1}{2} \langle{\rm C},H|.
\end{equation}
The weak value of the projection on $C$  at time $t_2$ is 
 \begin{equation}\label{wvA}
({\mathbf{P}_C})_w (t_2)=\frac{\langle \varphi (t_2)| {\mathbf{P}_C} |\psi (t_2)\rangle}{\langle \varphi (t_2)| \psi (t_2)\rangle}=\frac{1}{2}
.\end{equation}
Thus, the photon leaves the trace half the magnitude of a bouncing photon, so according to the definition in \cite{past}, the photon was near   $MR_B1$.

At time $t'_2$, given that the photon was not detected by $D_A1$, the forward evolving state is the same as at $t_2$, i.e. it is given by the right hand side of (\ref{psi}), but the backward evolving state is, instead of (\ref{phi}), $\langle \varphi(t'_2)|=\langle{\rm A},H|$. Thus, the weak value of the projection on $C$ is
\begin{equation}\label{wvA'}
({\mathbf{P}_C})_w (t'_2)=\frac{\langle \varphi (t'_2)| {\mathbf{P}_C} |\psi (t'_2)\rangle}{\langle \varphi (t'_2)| \psi (t'_2)\rangle}=0
.\end{equation}
Therefore, the photon was not present near the mirror $MR_B3$.

Salih's paradox appears when he tries to analyse the situation at an intermediate time $t_4$, after the event of not detection by the detector $D_A1$.   Apparently, the future action  affects the past! He argues that at time $t_4$ there is no trace at the mirror $MR_B1$. However, the construction of the second interferometer leads to a finite probability of a click at $D_0$, after which the TSVF asserts  the presence of the trace at the mirror $MR_B1$. So, it seems that creation of the second interferometer  with the detector $D_0$ leads (with non-vanishing probability) to the appearance of the trace in the past. Disappearance of the trace in the past for a probabilistic outcome in the future is not surprising, but creation of the trace which was not present there before, is a paradox.

Salih argues ``from within the weak measurement framework'' that the TSVF assertion is mistaken, that the photons were ``never at C'', in particular, there is no trace at $MR_B1$. Salih's resolution also restores the classical ``common sense'' picture according to which the photon reaching $D_0$ could not have been at C. We argue, however, that Salih's weak measurement analysis is incorrect. It is a subtle error because the analysis of this question in the  framework of the TSVF is not straightforward. We will perform it in Sec.~4, but first, to avoid the controversies related to the TSVF, we will present an analysis in the framework of the standard formalism without invoking backward evolving quantum state.

\section{Trace analysis without TSVF}

Describing the trace of a flying photon  is a difficult physics question, but the trace of a photon bouncing off a mirror  
is well defined. The mirror gets a finite momentum from the photon. In a good interferometer, this momentum is much smaller than the quantum uncertainty of the momentum of the mirror because the mirror has to be localised with the uncertainty smaller than the photon wavelength. So, we can describe the evolution of the quantum state of the mirror  $| \chi \rangle$ when a photon bounces off the mirror  as 
\begin{equation}\label{mirror}
|\chi \rangle \rightarrow|\chi^\prime \rangle \equiv \eta\left( |\chi \rangle + \epsilon |\chi^\perp \rangle \right),
\end{equation}
where $| \chi \rangle$ is the quantum state of the mirror when photon is not present and
$| \chi^\perp \rangle$ denotes the component of $| \chi^\prime \rangle$ which is orthogonal to $| \chi \rangle$.
By definition we choose the phase of $|\chi_1^\perp\rangle$ such that $\epsilon > 0$.
The trace left by the photon is manifested by the presence of the orthogonal component $| \chi^\perp \rangle$ and is quantified by the (small) parameter $\epsilon$.

Now we want to analyze the trace in the Salih interferometer at time $t_4$. In fact, there will be traces in all mirrors and beam splitters of the interferometer described by equations similar to (\ref{mirror}) and it might be fruitful to take all of them into account in order to obtain the complete picture of traces of the photon inside the interferometer. However, since we are interested in the trace on mirror $MR_B1$, we will consider only the quantum state of the photon and that mirror. Disregarding other optical components will make changes of the second order in $\epsilon$ regarding the trace on the mirror $MR_B1$, while only the trace of the first order in $\epsilon$ is of interest. (Later, considering the evolution in the second cycle, we will add the description of the mirror $MR_B3$.)

We introduce a set of new intermediate times, see Fig. 1,  and describe (neglecting terms of the second order in $\epsilon$) the time evolution of the state of the photon and the mirror at times 
$t_2$, $t_5$, $t_6$, $t_7$ and $t_8$:
\begin{eqnarray}\label{psi(t)}
\left (\frac{1}{\sqrt 2} |{\rm A},H\rangle + \frac{1}{2} |{\rm B},V\rangle-\frac{1}{2} |{\rm C},H\rangle \right ) | \chi_1 \rangle \rightarrow \\
\left (\frac{1}{\sqrt 2} |{\rm A},H\rangle  + \frac{1}{2} |{\rm B},V\rangle \right )| \chi_1 \rangle-\frac{1}{2} |{\rm C},H\rangle  | \chi_1^\prime\rangle \rightarrow  \\
\frac{1}{\sqrt 2} \left [ |{\rm A},H\rangle  + \frac{1}{2}( |{\rm D},V\rangle-|{\rm D},H\rangle) \right ]| \chi_1 \rangle-\frac{1}{2\sqrt 2}( |{\rm D},V\rangle+|{\rm D},H\rangle ) | \chi_1^\prime\rangle \rightarrow   \\
\frac{1}{\sqrt 2} \left( |{\rm S},H\rangle   - |{\rm J},H\rangle\right) |\chi_1 \rangle-\frac{\epsilon}{2\sqrt 2}(|{\rm S},V\rangle +|{\rm J},H\rangle) |\chi_1^\perp\rangle \rightarrow \\
 |{\rm S},H\rangle |\chi_1 \rangle -\frac{\epsilon}{2} |{\rm S},V\rangle| \chi_1^\perp\rangle.
\end{eqnarray}
The presence of the orthogonal component of the mirror state of the first order in $\epsilon$ tells us that (contrary to Salih's claim) the photon left a trace in $MR_B1$.
Continuing the time evolution in the second outer interferometer adding to the consideration the quantum state $| \chi_3 \rangle $ of the mirror $MR_B3$ for times  $t_8$, $t_9$,  $t_{10}$, and  $t_{11}$, yields:
\begin{eqnarray}\label{evo2}
 \left(|{\rm S},H\rangle |\chi_1 \rangle -\frac{\epsilon}{2} |{\rm S},V\rangle| \chi_1^\perp\rangle \right) |\chi_3\rangle \rightarrow \\
\frac{1}{\sqrt 2}\left [ ( |{\rm S},H\rangle  - |{\rm J},H\rangle)| \chi_3 \rangle-\frac{\epsilon}{2}(|{\rm S},V\rangle +|{\rm J},H\rangle)| \chi_3^\perp\rangle \right ]| \chi_1 \rangle +\nonumber \\ \frac{\epsilon}{2\sqrt 2} ( |{\rm S},H\rangle  + |{\rm J},H\rangle)|  \chi_3\rangle \chi_1^\perp \rangle
\rightarrow \\
|{\rm F},H\rangle |\chi_3 \rangle \left(| \chi_1\rangle+ \frac{\epsilon}{2} | \chi_1^\perp\rangle\right )-\frac{\epsilon}{2}|{\rm G},V\rangle| \chi_3^\perp\rangle | \chi_1 \rangle
\rightarrow \\
|{\rm F},H\rangle |\chi_3 \rangle \left(| \chi_1\rangle+ \frac{\epsilon}{2} | \chi_1^\perp\rangle\right ).
\end{eqnarray}
We see that the trace in $MR_B1$ remains after detection by $D_0$.

The second cycle has an additional polarization beam splitter that filters the $H$ polarization for photons in F. Without it, as it can be seen from (13), there are traces both in $MR_B1$ and in $MR_B3$, so the paradox of the difference of traces in identical cycles does not arise. This explains also Salih's sentence: ``Our resolution of the paradox,  supported
by recent weak-measurement {\it experimental results} \cite{SalihNPJ}, shows that the photon has not been to
any of the right-hand-side mirrors.'' The experiment \cite{SalihNPJ} was done with the additional beam splitter observing the signal on photons detected by $D_0$, see Fig.~2.  So, the experiment implemented the second (and not the first) outer cycle that has an additional beam splitter and thus, indeed, has no trace in C. 

\begin{figure} [ht]
   \begin{center}
     \includegraphics[width=12cm]{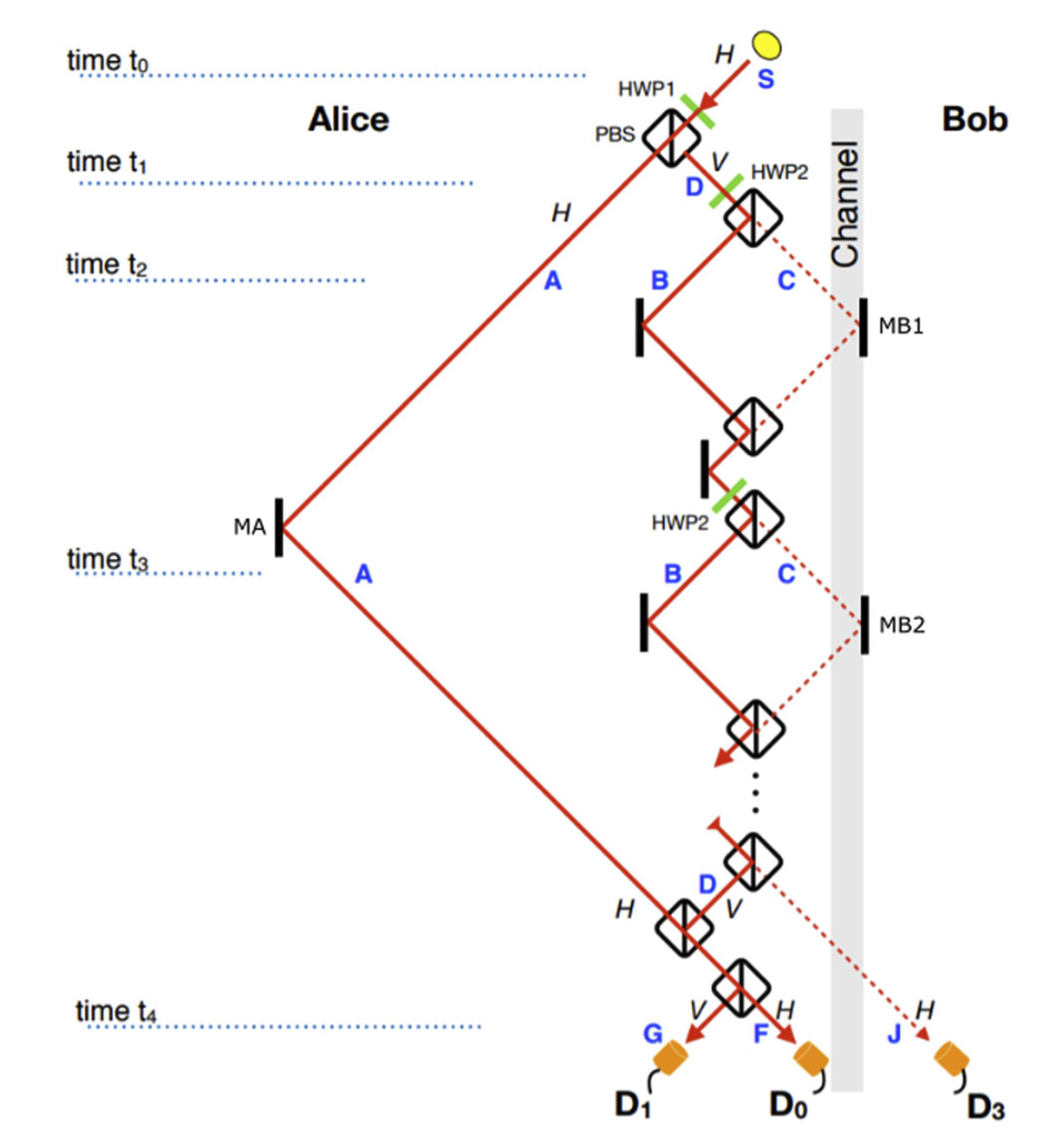}
     \end{center}
   \caption
   { \label{fig1}
   {\bf Experimentally implemented ``one-cycle counterfactual communication protocol''}. (Fig.~1 of \cite{SalihNPJ}.) 
The experiment demonstrates that photons detected by $D_0$ leave no trace on mirrors $MBi$.  Salih views these results  as a support of his resolution of the paradox  ``that the photon has not been to
any of the right-hand-side mirrors''.
However, this setup is equivalent only to the second cycle of the experiment of Fig.~1 which has an additional polarisation beam splitter  and for which the TSVF predicts no traces on the right-hand-side mirrors. The experiment can shed no light on the traces in the first cycle of Fig.~1.
}
   \end{figure}

Note, that the  counterfactuality in the setup \cite{SalihNPJ} presented in Fig. 2 is very limited. The protocol sometimes tests the presence of the Bob's shutter using the Zeno modification of the interaction-free measurement (first proposed in \cite{Ho06}). It is counterfactual when the shutter is present (and  $D_1$ clicks), but not counterfactual when  the shutter is absent and   $D_3$ clicks. Or, sometimes, it does not test the presence of the shutter and   $D_0$ clicks. Indeed, Alice can know in a counterfactual  way only about the presence of the shutter (when $D_1$ clicks). The other click ($D_0$) only tells her that the presence of the shutter was not tested.
The double cycle protocol of Fig.~1 is also not fully counterfactual, but fortunately, it is conceptually different from  Salih's counterportation procedure.
 
\break

\section{The analysis within TSVF}

Now we are in the position to discuss Salih's  TSVF analysis. Salih wrote:
\begin{quote}
Our resolution of the paradox, from within the weak measurement framework, is based on
the observation that the strong measurement by detector $D_A$ at the end of each outer cycle
projects the state of the photon onto arm S, where we know it should be $H$-polarised. This is
the post-selected state.
 \end{quote}
The main Salih's error is the sentence: ``This is
the post-selected state.'' He uses it as a backward evolving state of the TSVF. It is not, it is the forward evolving state of the TSVF at $t_4$. ``The strong measurement by detector $D_A$'' is not a complete measurement, so we do not have the backward evolving quantum state and, consequently, the TSVF cannot be applied in a straightforward way.

A possible strategy of applying the TSVF in such a case  is to use the fact that the future does not affect the past
and consider some (hypothetical) complete measurement on the system in the future, see \cite{beyond}. There are several options. We will consider three options in the following three paragraphs and will see that the trace is the same independently of the future measurement.

We have calculated the forward evolving state after the nondetection by $D_A$. It is a quantum state of the composite system of the photon and the mirror (9). Verification measurement of this state, if performed, will succeed with certainty, so we will end up with a complete two-state vector description which can provide the trace we want to know. However, it requires the analysis of the composite system: the simple consideration of the weak value of the projection of the photon on the mirror location does not explain the trace. Indeed, the calculation yields
$({\mathbf{P}_C})_w(t_2)=0$, so the trace can be found only by the direct analysis of the mirror variables, e.g., its momentum.  So, this method is consistent, but not very useful. The advantage of the TSVF is in simplifying the description by limiting it to the description of the system itself, without (as it required in the standard formalism) involving the external systems it interacts with. 

We can also consider measurement of the system only, e.g., the polarisation measurement of the photon exiting the first cycle towards S. This, in fact, is done by Salih thorough introducing the second cycle. Detection by $D_A2$ corresponds to finding 
$\frac{1}{\sqrt 2}  ( |H\rangle+|V\rangle)$, while detection by detection by $D_0$
corresponds to finding $\frac{1}{\sqrt 2}  ( |H\rangle-|V\rangle)$. 
The standard TSVF procedure for complete postselection at $D_0$ yields 
$({\mathbf{P}_C})_w (t_2)=\frac{1}{2}$ and significant trace on the mirror is  manifested in  the appearance of the component 
$\frac{\epsilon}{2}|\chi_1^\perp\rangle$ corresponding to the momentum kick of the mirror. Detection by 
$D_A2$ leads to $({\mathbf{P}_C})_w (t_2)=-\frac{1}{2}$ and the appearance of the component 
$-\frac{\epsilon}{2}|\chi_1^\perp\rangle$ corresponding to the  momentum kick of the mirror in the opposite direction. The probability of defections by $D_A2$ and $D_0$ are  equal, so the expectation value of the momentum of the mirror remains as before. This explains $({\mathbf{P}_C})_w(t_2)=0$ for the mixed postselected state. (The weak value in the case of mixed states has been derived in \cite{beyond}, Eq.~(32) therein.) Still, the magnitude of the trace on the mirror in the mixed state of equal kicks in the two directions as the change of the undisturbed quantum state is not less than the change due to a single kick. Thus, placing detector $D_0$ does not lead to increase of the trace in the past.

One can also consider a simple measurement in the $H, V$
basis. 
With the probability close to 1, the result is $H$, in which case, as Salih correctly observes, there is no trace on the mirror $MR_B1$. But with probability $\frac{\epsilon^2}{4}$  we get $V$. In this case, the trace is anomalously large, it is not of the order $\epsilon$, the mirror state is orthogonal to the undisturbed state. So, considering the mixture of the results, we obtain the same trace again.

\section{Conclusions}

Salih's alleged paradox according to which the evolution in the two cycles is identical, but in one of them the photon kicks the mirror and in other does not, is naturally resolved within the TSVF which asserts that  the  complete description of a quantum pre and postselected system is given by the two-state vector which includes, in addition to the standard, forward evolving wave function, the backward evolving quantum state. The  trace differences in the two cycles are explained by the fact that the backward evolving states at $t_2$ and $t'_2$ are different.

Salih's ``paradox'' demonstrates once more that classical ``common sense'' is not applicable for quantum particles. Photon, as a classical particle with a continuous trajectory, could not have been near the mirror the quantum state of which it changed. (Similar situation was demmonstrated in \cite{Danan}.)

Salih's  paradox demonstrates the subtle issues  of the TSVF. When the postselection measurements are not complete,   we cannot neglect the weak interaction with the environment, the interaction which is usually neglected in the weak value formulae of the TSVF. When the measurements are not complete, the weak values of the system's variables do not faithfully describe the weak trace. In mixed states weak values faithfully describe the expectation value of the weak measurement pointer position, but not the other aspects of quantum state of the pointer, see \cite{beyond}. When the postselection is not complete, the TSVF approach is consistent, but it does not necessarily provide an advantage of simplicity as in the case of complete pre and postselection measurements.

 Salih's error in the analysis of  his ``paradox''  does not invalidate his interesting proposal for counterportation: a transfer of a quantum state in such a way that the carriers of the information are not present in the transmission channel (they do not leave a trace of the same order of magnitude as a localised carrier would leave). However, it shows that the method holds only if the modification \cite{AV19} is applied.

This work has been supported in part by the U.S.-Israel Binational Science Foundation (Grant No. 735/18). JD was partially supported by National Science Foundation (NSF) (Grant No. 1915015) and Army Research Office (ARO) (Grant No. W911NF-18-1-0178).

\bibliographystyle{unsrt}

\end{document}